\documentstyle[prd,aps,eqsecnum,preprint,tighten,floats,epsf]{revtex}

\begin{document}

%remove for camera-ready copy
\preprint{\vbox{\hfill UMN-D-99-3  \\
          \vbox{\hfill September 1999 } 
          \vbox{\vskip0.3in}}}

\title{On the Use of Discrete Light-Cone Quantization \\
to Compute Form Factors%
%remove footnote for camera-ready copy
\footnote{To appear in the proceedings of the TJNAF Workshop
on The Transition from Low to High $Q$ Form Factors,
Athens, GA, September 17, 1999.}
}

\author{John R. Hiller}
\address{Department of Physics,
University of Minnesota, Duluth, Minnesota 55812}

\maketitle

\begin{abstract}
Techniques for the field-theoretic calculation of a form factor
are described and applied to a dressed-fermion state of a 
(3+1)-dimensional model Hamiltonian.  Discrete light-cone
quantization plays the crucial role as the means by which
Fock-state wave functions are computed.  An ultraviolet
infinity is controlled by Pauli--Villars regularization.
\end{abstract}

\section{Introduction}

There has been assembled a sequence of technologies by
which one might eventually compute hadronic form factors
directly from quantum chromodynamics (QCD).  These include
the early work by Drell and Yan \cite{DrellYan} and by
Brodsky and Drell \cite{BrodskyDrell} on the relation of
form factors to Fock-state wave functions in light-cone
quantization.  The wave functions can be calculated,
in principle, by the method of discrete light-cone
quantization (DLCQ) proposed by Pauli and Brodsky \cite{PauliBrodsky}.
Refinements of DLCQ that permit substantive calculations
in non-super-renormalizable, (3+1)-dimensional field theories
have now been tested by Brodsky, Hiller, and McCartor
\cite{PV1,PV2}; a key role is played by Pauli--Villars
(PV) regularization \cite{PauliVillars}, implemented
through the introduction of PV bosons to the DLCQ Fock-state
basis.  In the following sections, a description is given
of how this sequence comes together, and the steps are
applied to a form factor calculation in the model
of Ref.~\cite{PV2}.

\section{Form Factors and Wave Functions}

Formal expressions for the Dirac and Pauli
form factors $F_1$ and $F_2$ of a composite fermion
can be obtained \cite{DrellYan,BrodskyDrell}
by considering the current matrix elements
\begin{eqnarray}
\langle P+Q,\sigma | J^+(0)/P^+ |P,\sigma\rangle&=&2F_1(Q^2)\,,
\\ \nonumber
\langle P+Q,\sigma | J^+(0)/P^+ |P,-\sigma\rangle
   &=&-4\sigma(Q_x-iQ_y)F_2(Q^2)/2M\,,
\end{eqnarray}
where $J^+=J_0+J_z$ is the plus component of the current, $P^+=E+P_z$ is
the light-cone longitudinal momentum of the initial state of mass $M$,
$Q$ is the photon momentum, and $\sigma$ is the 
spin projection along the $z$ axis.  By working in light-cone coordinates
\cite{Dirac,DLCQreview} and in the Drell-Yan frame \cite{DrellYan},
with $Q=(0,2Q\cdot P/P^+,\bbox{Q}_\perp)$, the form factors can
be expressed directly in terms of Fock-sector wave functions
$\psi^{(n)}$ as \cite{BrodskyDrell}
\begin{eqnarray} \label{eq:F1}
F_1(Q^2)&=\sum_n\sum_je_j&\int\delta(1-\sum_i x_i)\prod_i dx_i \nonumber \\ 
&&  \times 16\pi^3\delta(\sum_i \bbox{k}_{\perp i})
      \prod_i \frac{d^2k_{\perp i}}{16\pi^3}
   \psi_{P+Q,1/2}^{(n)*}(x,\bbox{k}_\perp^\prime)
   \psi_{P,1/2}^{(n)}(x,\bbox{k}_\perp)\,, \nonumber  \\ 
-\left(\frac{Q_x-iQ_y}{2M}\right) F_2(Q^2)&=
   \sum_n\sum_je_j&\int\delta(1-\sum_i x_i)\prod_i dx_i  \\
&&  \times 16\pi^3\delta(\sum_i \bbox{k}_{\perp i})
        \prod_i \frac{d^2k_{\perp i}}{16\pi^3}
   \psi_{P+Q,1/2}^{(n)*}(x,\bbox{k}_\perp^\prime)
   \psi_{P,-1/2}^{(n)}(x,\bbox{k}_\perp)\,.  \nonumber
\end{eqnarray}
Here $n$ is the number of constituents; $e_j$ is the charge of the
struck constituent; the $x_i$ are longitudinal momentum fractions
$p_i^+/P^+$ for constituents with momenta $p_i$; 
the $\bbox{k}_{\perp i}$ are relative transverse momenta; and
\begin{equation}
\bbox{k}_{\perp i}^\prime=\left\{\begin{array}{ll}
       \bbox{k}_{\perp i}-x_i\bbox{Q}_\perp\,, & i\neq j \\
       \bbox{k}_{\perp j}+(1-x_j)\bbox{Q}_\perp\,, & i=j \end{array}\right.
\end{equation}
are transverse momenta relative to the new $P+Q$ direction.
Thus, given the wave functions $\psi^{(n)}$ one can compute the form
factors.

In light-cone quantization these wave functions can be found by 
diagonalizing the mass-squared operator $P^+P^--P_\perp^2\equiv H_{\rm LC}$,
traditionally called the light-cone Hamiltonian \cite{DLCQreview}.  
This is simplest
in a frame where the total transverse momentum is zero and in a 
basis where $P^+$ is diagonal.  Then only the light-cone energy 
operator $P^-$ has a nontrivial representation, and the eigenstate
can be expanded explicitly in terms of momentum Fock states
$|n:x_i\,,\,\bbox{k}_{\perp i} \rangle$ and the desired
wave functions:
\begin{equation}
|P,\sigma\rangle=\sum_n\int\delta(1-\sum_i x_i)\prod_i \frac{dx_i}{\sqrt{x_i}}
  16\pi^3\delta(\sum_i \bbox{k}_{\perp i})\prod_i \frac{d^2k_{\perp i}}{16\pi^3}
   \psi_{P,\sigma}^{(n)}(x,\bbox{k}_\perp)
     |n:x_i\,,\,\bbox{k}_{\perp i} \rangle\,.
\end{equation}
It is a significant advantage of light-cone coordinates that such an
expansion is well-defined, in the sense that there are no contributions from
disconnected vacuum pieces \cite{DLCQreview}.  Another advantage is that
the boost invariance of $x$ and $\bbox{k}_\perp$ permit the same wave
functions to be used in the construction of the boosted state
$|P+Q,\sigma\rangle$.

\section{Computation of Wave Functions}

The wave functions can be computed by applying DLCQ
\cite{PauliBrodsky,DLCQreview} to the eigenvalue problem 
$H_{\rm LC}|P,\sigma\rangle=M^2|P,\sigma\rangle$.
Pauli and Brodsky applied this method to (1+1)-dimensional 
models for its initial trials \cite{PauliBrodsky}.  Since then
much work has been done \cite{DLCQreview}, particularly in
1+1 dimensions.  The greater complexity of (3+1)-dimensional
theories has made progress there much slower.  The need
for regularization and nonperturbative renormalization 
has been most telling \cite{Wilson}.

Recently a scheme for use of Pauli--Villars (PV) ultraviolet
regularization \cite{PauliVillars} within DLCQ calculations
has been successfully tested for (3+1)-dimensional model 
Hamiltonians \cite{PV1,PV2}.  Massive PV bosons are introduced 
as additional constituents in the Fock basis, with imaginary 
couplings chosen to produce desired cancellations in perturbation 
theory.  The eigenvalue problem is solved nonperturbatively,
and the limit of infinite PV mass is taken.

Of the two model Hamiltonians considered, the more
sophisticated is \cite{PV2}
\begin{eqnarray}
H_{\rm LC}&=&\int\frac{dp^+d^2p_\perp}{16\pi^3p^+}
                                  (\frac{M^2+p_\perp^2}{p^+/P^+}+M'_0p^+/P^+)
                 \sum_\sigma b_{\underline{p}\sigma}^\dagger b_{\underline{p}\sigma} 
\nonumber \\
  & & +\int\frac{dq^+d^2q_\perp}{16\pi^3q^+}
       \left[\frac{\mu^2+q_\perp^2}{q^+/P^+}a_{\underline{q}}^\dagger a_{\underline{q}}
           + \frac{\mu_1^2+q_\perp^2}{q^+/P^+}a_{1\underline{q}}^\dagger a_{1\underline{q}}
               \right]
\nonumber  \\
   &  & +g\int\frac{dp_1^+d^2p_{\perp1}}{\sqrt{16\pi^3p_1^+}}
            \int\frac{dp_2^+d^2p_{\perp2}}{\sqrt{16\pi^3p_2^+}}
              \int\frac{dq^+d^2q_\perp}{16\pi^3q^+}
                \sum_\sigma b_{\underline{p}_1\sigma}^\dagger b_{\underline{p}_2\sigma}
   \\
    &  & \rule{0.25in}{0mm}\times \left[
         a_{\underline{q}}^\dagger\delta(\underline{p}_1-\underline{p}_2+\underline{q})
        +a_{\underline{q}}\delta(\underline{p}_1-\underline{p}_2-\underline{q}) \right.
\nonumber \\
    &  & \rule{0.5in}{0mm} \left.
     +i a_{1\underline{q}}^\dagger\delta(\underline{p}_1-\underline{p}_2+\underline{q})
    +i a_{1\underline{q}}\delta(\underline{p}_1-\underline{p}_2-\underline{q}) \right]\,.
\nonumber 
\end{eqnarray}
Fermions created by $b_{\underline{p}\sigma}^\dagger$, with
light-cone momentum $\underline{p}\equiv(p^+,p_x,p_y)$ and spin $\sigma$,
act as sources and sinks for physical and PV bosons created 
by $a_{\underline{q}}^\dagger$ and $a_{1\underline{q}}^\dagger$.
The fermion mass is $M$, and the boson masses $\mu$
and $\mu_1$.  The imaginary coupling of the PV boson
causes a cancellation of an infinity in the
fermion self-energy.  The $M'_0p^+/P^+$ counterterm is then
all that is needed to remove the shift in the mass,
with $M'_0\sim\ln\mu_1/\mu$.

The eigenvalue problem for this Hamiltonian, in the one-fermion
sector with total momentum $\underline{P}$, reduces to a 
system of integral equations
\begin{eqnarray} \label{eq:CoupledEqns}
\lefteqn{\left[M^2-\frac{M^2+p_\perp^2}{p^+/P^+}-M'_0p^+/P^+
  -\sum_i\frac{\mu^2+q_{\perp i}^2}{q_i^+/P^+}
                  -\sum_j\frac{\mu_1^2+r_{\perp j}^2}{r_j/P^+}\right]
                    \psi^{(n,n_1)}(\underline{q}_i,
                                       \underline{r}_j,\underline{p}) } 
\nonumber \\
& & =g\left\{\sqrt{n+1}\int\frac{dq^+d^2q_\perp}{\sqrt{16\pi^3q^+}}
              \psi^{(n+1,n_1)}(\underline{q}_i,\underline{q},
                              \underline{r}_j,\underline{p}-\underline{q})\right.
\\
& &\rule{0.5in}{0mm} +\frac{1}{\sqrt{n}}\sum_i\frac{1}{\sqrt{16\pi^3q_i^+}}
       \psi^{(n-1,n_1)}(\underline{q}_1,\ldots,\underline{q}_{i-1},
                                 \underline{q}_{i+1},\ldots,\underline{q}_n,
                                  \underline{r}_j,\underline{p}+\underline{q}_i)
\nonumber \\
& &\rule{0.5in}{0mm}+i\sqrt{n_1+1}\int\frac{dr^+d^2r_\perp}{\sqrt{16\pi^3r^+}}
              \psi^{(n,n_1+1)}(\underline{q}_i,\underline{r}_j,
                                     \underline{r},\underline{p}-\underline{r})
\nonumber \\
& &\rule{0.5in}{0mm}+\left.\frac{i}{\sqrt{n_1}}\sum_j\frac{1}{\sqrt{16\pi^3r_j^+}}
 \psi^{(n,n_1-1)}(\underline{q}_i,\underline{r}_1,\ldots,\underline{r}_{j-1},
                                 \underline{r}_{j+1},\ldots,\underline{r}_{n_1},
                                     \underline{p}+\underline{r}_j) \right\}
\nonumber 
\end{eqnarray}
for the wave functions $\psi^{(n,n_1)}$, where $n$ and $n_1$ are the
numbers of physical and PV bosons.  There are two bare parameters
$M'_0$ and $g$, which are fixed by setting values for physical
quantities chosen to be $M$ and
$\langle :\!\!\phi^2(0)\!\!:\rangle
   \equiv\langle P,\sigma|\!:\!\!\phi^2(0)\!\!:\!|P,\sigma\rangle$.
The latter was chosen for ease of computation, in the form
\begin{eqnarray}
\langle :\!\!\phi^2(0)\!\!:\rangle
        &=&\sum_{n=1,n_1=0}^\infty\prod_i^n\int\,dq_i^+d^2q_{\perp i} 
                \prod_j^{n_1}\int\,dr_j^+d^2r_{\perp j}
   \\
    & & \rule{0.25in}{0mm} \times \left(\sum_{k=1}^n \frac{2}{q_k^+/P^+}\right)
              \left|\psi^{(n,n_1)}(\underline{q}_i,\underline{r}_j;
       \underline{P}-\sum_i\underline{q}_i-\sum_j\underline{r}_j)\right|^2\,.
\nonumber
\end{eqnarray}
These renormalization conditions must be solved simultaneously with
the eigenvalue problem.

The DLCQ method translates the field-theoretic eigenvalue problem
into a matrix problem.  Discrete momentum values $p^+=n\pi/L$
and $\bbox{p}_\perp=(n_x\pi/L_\perp,n_y\pi/L_\perp)$ are used,
with $L$ and $L_\perp$ the length scales associated with the approximation.
Momentum integrals are approximated by trapezoidal sums over the
discrete points.  Near the endpoints special weighting factors are 
needed for better accuracy, as discussed in Ref.~\cite{PV1}.

The length scales $L$ and $L_\perp$ define a coordinate-space box
within which the fields are assigned periodic boundary conditions,
except for a longitudinal antiperiodic condition for the fermion.
The integer $n$ is then odd for fermions but even for bosons.
The total longitudinal momentum for a one-fermion state
defines an (odd) integer $K=P^+L/\pi$ called the harmonic
resolution \cite{PauliBrodsky}.  The transverse integers
$n_x$ and $n_y$ range between $-N_\perp$ and $N_\perp$, with
$N_\perp$ fixed by a cutoff $m_i^2+p_{\perp i}^2<\Lambda^2p_i^+/P^+$.
The cutoff is needed to produce a matrix of finite size but
is not used as a regulator.  The parameters $K$, $L_\perp$,
and $\Lambda^2$ determine the numerical approximation.

The matrix eigenvalue problem is readily solved for the lowest
massive state by use of the complex-symmetric Lanczos
algorithm \cite{Lanczos}.  The algorithm allows the matrix
to be stored in a compact form and to be referenced only in
matrix-vector multiplications.  This feature, combined
with the extreme sparseness of the matrix, has permitted
calculations with as many as 10.5 million basis states \cite{PV2}.

\section{Computation of a Form Factor}

With these techniques in place we may consider calculation
of $F_1$ for the dressed fermion state of the model.\footnote{The
interactions of the model do not flip the spin, and $F_2$ is therefore
identically zero.}  In Ref.~\cite{PV2} only the slope $F'_1(0)$
was computed, from an expression derived from (\ref{eq:F1}).
Here we compute with (\ref{eq:F1}) directly.
 
\begin{figure}[h]
\centerline{\epsfxsize=\columnwidth \epsfbox{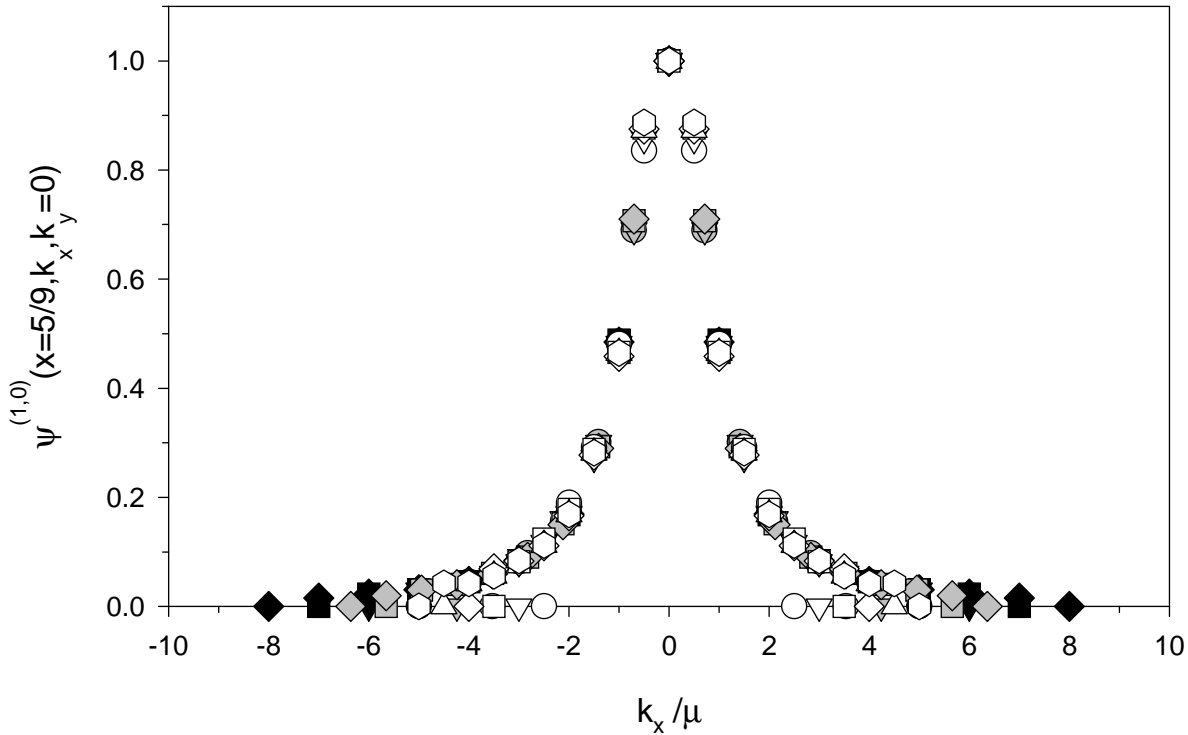} }
\caption{\label{fig:psi10}
A cross section of the boson-fermion two-body amplitude
taken at fixed longitudinal momentum fraction $x=5/9$
and at fixed $k_y=0$, with $K=9$,
$\langle:\!\!\phi^2(0)\!\!:\rangle=1$,
and $\mu_1^2=10\mu^2$.  The cutoff $\Lambda^2$ and
the transverse resolution $N_\perp$ are varied to
keep the transverse scale $L_\perp$ fixed
at one of the following values: $1\frac{\pi}{\mu}$ (black),
$\sqrt{2}\frac{\pi}{\mu}$ (gray),
and $2\frac{\pi}{\mu}$ (white).  Different symbol shapes
correspond to different values of $N_\perp$.
The peaks are normalized to be equal at $k_x=0$.
The points at zero amplitude mark the transverse range, which is
set by the cutoff.}
\end{figure}

The structure of the $F_1$ formula is that of an overlap
of momentum wave functions, with one shifted in a
transverse direction, which we take to be $x$ so that
$\bbox{Q}_\perp=Q_\perp\hat{x}$.  The shape of the
one-boson wave function is shown in Fig.~\ref{fig:psi10}.
For the coupling strength considered, this wave function
provides the primary contribution to the variable part
of $F_1$.  Of course, the bare fermion Fock state
provides a $Q_\perp$-independent contribution equal
to that state's probability.

\begin{figure}[h]
\centerline{\epsfxsize=\columnwidth \epsfbox{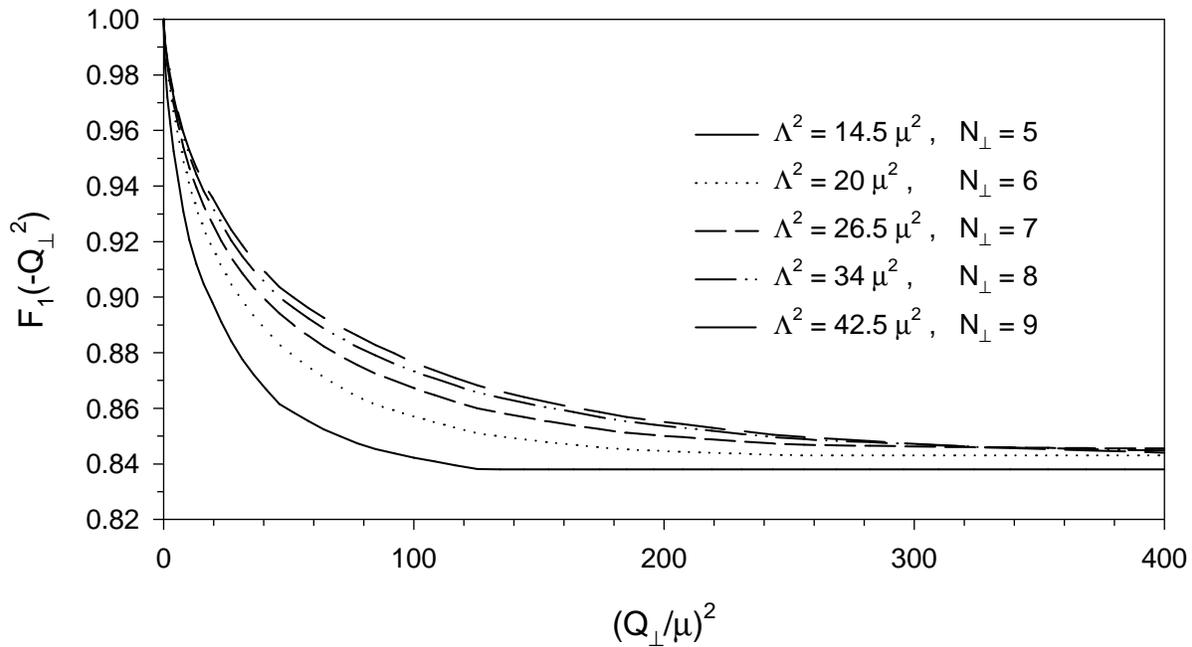} }
\caption{\label{fig:lambda}
The form factor $F_1$ for fixed resolution $K=9$ and $L_\perp=2\pi/\mu$.
Various cutoffs $\Lambda^2$ are considered.  The model parameter values are
$\langle:\!\!\phi^2(0)\!\!:\rangle=1$,
$M^2=\mu^2$, and $\mu_1^2=10\mu^2$. }
\end{figure}

\begin{figure}[h]
\centerline{\epsfxsize=\columnwidth \epsfbox{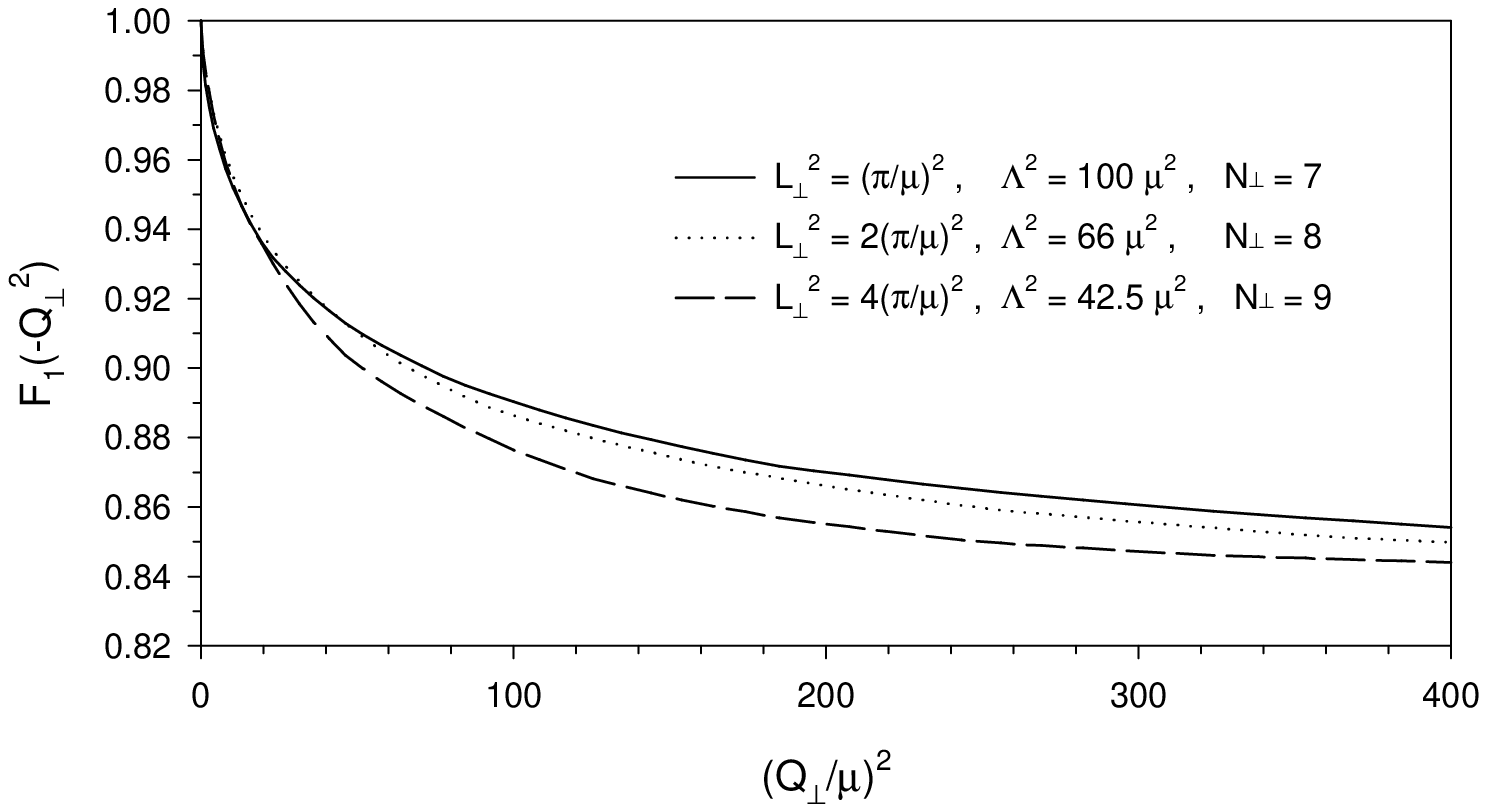} }
\caption{\label{fig:Lperp}
The form factor $F_1$ for fixed longitudinal resolution $K=9$ 
and varying transverse resolution $L_\perp$.
The largest available cutoff was used in each case.  
The model parameter values are
$\langle:\!\!\phi^2(0)\!\!:\rangle=1$,
$M^2=\mu^2$, and $\mu_1^2=10\mu^2$. }
\end{figure}

Calculations \cite{PV2} have shown that the wave 
functions quickly become independent of $K$ and $L_\perp$
as these parameters are increased.  The independence with
respect to $L_\perp$ can be seen in Fig.~\ref{fig:psi10}.
However, $F'(0)$ was
found to be sensitive to $L_\perp$, and one would 
expect the tail of $F_1$ to be sensitive to $\Lambda^2$ as
well as $L_\perp$.  That this is the case 
can be seen in Figs.~\ref{fig:lambda} and \ref{fig:Lperp}.
For small $\Lambda^2$ the form factor quickly reaches the
bare-fermion contribution.  For larger $\Lambda^2$ the
bare-fermion contribution increases slightly, and the
approach of $F_1$ to this limiting value becomes more
gradual.  When $\Lambda^2$ is large enough for the shape
of $F_1$ to appear converged there remains a significant
$L_\perp$ dependence, as seen in Fig.~\ref{fig:Lperp},
even though the one-boson wave function changes little with $L_\perp$.
The form factor remains sensitive because $L_\perp$ controls
the approximation to the integral of the wave function
product from which $F_1$ is computed.

\section{Summary}

These results show that, at least for model Hamiltonians, a
field-theoretic calculation of Fock-state wave functions
and bound-state form factors can be carried out.  The
added Pauli--Villars particles provide the ultraviolet 
regularization without making the basis size
unmanageable.  Work on a more complete field theory,
a single-fermion truncation of Yukawa theory, is
underway.  Consideration of the dressed-electron
and positronium states of quantum electrodynamics
would be a natural next step.  QCD will require a more
sophisticated approach, perhaps relying on heavy
supersymmetric partners to play the role of the
Pauli--Villars particles.

\section*{Acknowledgments}
This work is an extension of work done 
in collaboration with S.J. Brodsky and G. McCartor
and was supported in part by the Minnesota Supercomputing Institute
through grants of computing time and by the Department of Energy
contract DE-FG02-98ER41087.

\end{document}